# Eulerian Gaussian beams for high frequency wave propagation in inhomogeneous media of arbitrary anisotropy

Xingguo Huang

University of Bergen, Department of Earth Sciences, Allegaten 41, 5020 Bergen, Norway.

Email: xingguo.huang19@gmail.com.

## Abstract

We present the Eulerian Gaussian beam method in anisotropic media. We derive kinematic and dynamic ray tracing equations based on the level set theory and Eulerian theory using the anisotropic eikonal equation. Compared with the traditional anisotropic Gaussian beam method using ray-centered coordinates, the anisotropic Eulerian Gaussian beam method derived in this work has the following three advantages: (1) it can handle the problem of calculating the distance from the imaging point to the beam point more easily; (2) it allows the travel time and amplitude to be distributed uniformly within the actual computational domain without interpolation; (3) it can handle late arrivals, both theoretically and in calculations, due entirely to ray tracing in the phase space.

**Keywords:** Eulerian Gaussian beam; level set method; seismic anisotropy; wave propagation

## 1. Introduction

The Gaussian beam method is an extension of ray theory. As a result of its use of complex travel times and amplitudes, the Gaussian beam method can deal with the problems that the high-frequency ray theory faces with respect to singularities. The Gaussian beam method has been used for high-frequency wave field calculations (Červený, 2001; Červený et al., 1982; Popov, 1982) as well as for migration imaging (Hill, 1990, 2001; Gray, 2005; Gray and Bleistein, 2009). Much work has been carried out to improve the Gaussian beam method (Alkhalifah, 1995; Červený, 1985; George et al., 1987; Klimes, 1984; Kravtsov and Berczynski, 2007; Nowack, 2003; Nowack and Kainkaryam 2011; Popov et al., 2010;

Tanushev, 2008; Zhu et al., 2007). However, the beam is calculated using the ray-centered coordinate system and it is necessary to calculate the distance of the imaging point to the nearest point on the ray or to carry out coordinate transformation (Červený and Pšenčik, 2010; Klimes, 1994). This is a very time-consuming calculation. In contrast, the multi-value travel time is automatically included in the Gaussian beam method based on complex travel times, although a quantitative theoretical proof on late arrivals processed by the Gaussian beam method has not yet been reported. For these reasons, Leung et al. (2007) proposed the Eulerian Gaussian beam method.

The Eulerian Gaussian beam method is a high-frequency asymptotic expression of the wave field based on level set theory (Qian and Leung, 2004, 2006; Leung et al., 2004) and is established in the Cartesian coordinate system. To calculate the Eulerian Gaussian beam, ray tracing and dynamic ray tracing are performed and the partial differential equations are solved. However, unlike traditional dynamic ray tracing in the Gaussian beam method using ray-centered coordinates, ray tracing and dynamic ray tracing in the Eulerian Gaussian beam method is based on the level set theory and the eikonal equation in Eulerian theory. Ray tracing of a single ray in the coordinate system is conducted point by point, so there are additional difficulties resulting from the uniform distribution of the phase and amplitude in the entire space and interpolation is required to determine the amplitude. However, due that Eulerian Gaussian beam theory is developed in the Cartesian coordinate system, the calculation is performed on regular grid points. Thus, the calculated traveltime and amplitude are distributed uniformly.

In geophysics, to overcome the shortcomings of the ray method in calculating the travel time, eikonal calculations based on the finite difference method (Vidale, 1990; van Trier and Symes, 1991; Sethian, 1996) are widely used. The method is based on the Eulerian theory, but only to calculate the first arrival – that is, the minimum phase – not to calculate later arrivals. Calculation at the level set theory has therefore been used in this work.

The aim of the work reported here was to obtain the Eulerian Gaussian beam in anisotropic media. Based on the anisotropic eikonal equation, we determined the derivative formula for ray tracing and then substituted it into the kinetic ray tracing equations to obtain the anisotropic dynamic ray tracing equations. We also modified the phase formula and obtained the anisotropic travel time formula. Finally, we derived weighting functions and superposition formulas for the Eulerian Gaussian beam based on the steepest descent and the stationary phase theory in anisotropic media. As a result of space constraints, this discussion is limited to the theory and there is no discussion and presentation of the numerical methods.

We review elastodynamic theory and some basic concepts of high-frequency wave fields. And then we deduce the basic formulas required in ray tracing from the anisotropic eikonal equation to obtain the ray tracing and dynamic ray tracing equations for anisotropic Eulerian Gaussian beam. Finally, we determine the anisotropic Eulerian Gaussian beam and Green's function by using Eulerian Gaussian beam superposition.

## 2. Basic concepts and equations

The elastic dynamic equation in anisotropic media (Červený, 2001) is as follows

$$\left(c_{ijkl}u_{k,l}\right)_{,j} = \rho \ddot{u}_i \tag{1}$$

where $u_i$ is the wave field, $u_{k,l}$ are the Cartesian components of the displacement vector $c_{ijkl}$ is the elasticity modulus and ρ is the density.

The high-frequency seismic wave field can be expressed as (Červený, 2001)

$$u(x_i, t) = U(x_i) \exp\{-i\omega[t - T(x_i)]\} \tag{2}$$

where $U$ represents the amplitude. To find the amplitude of the ray, it is necessary to know the initial amplitude at the starting point and the corresponding geometrical spreading factor. $T$ is the travel time of the seismic wave. Substituting this into Equation (1), we obtain the following equation (Červený and Pšenčik, 2010)

$$\left(\Gamma_{ik} - \delta_{ik}\right)U_k = 0 \tag{3}$$

The Christoffel matrix $\Gamma_{ik}(x_m, p_n)$ can be expressed as (Červený and Pšenčik, 2010)

$$\Gamma_{ik}(x_m, p_n) = a_{ijkl} p_j p_l \tag{4}$$

In this formula, $p_j$, $p_l$ are the slowness parameters and the elastic modulus is (Červený and Pšenčik, 2010)

$$a_{ijkl}(x_n) = \frac{c_{ijkl}(x_n)}{\rho(x_n)} \tag{5}$$

## 3. Paraxial anisotropic eikonal equation

Considering the Hamiltonian form of the eikonal equation in two-dimensional anisotropic media (Qian and Leung, 2004):

$$F(x, z, p_1, p_3) = 0 \tag{6}$$

where the slowness parameters $p_1, p_3$ are equal to:

$$\begin{aligned} p_1 &= \frac{\sin\theta}{v(x,z,\theta)} \\ p_3 &= \frac{\cos\theta}{v(x,z,\theta)} \end{aligned} \tag{7}$$

$\theta$ is the phase angle and $v$ is the wave velocity.

To obtain the ray tracing system, we give the relationship between dx, dz and $\tau$. According to Qian and Leung's (2004) research, we obtain

$$\frac{dx}{d\tau} = (p_1 \frac{\partial F}{\partial p_1} + p_3 \frac{\partial F}{\partial p_3})^{-1} \frac{\partial F}{\partial p_1} \tag{8}$$

$$\frac{dz}{d\tau} = (p_1 \frac{\partial F}{\partial p_1} + p_3 \frac{\partial F}{\partial p_3})^{-1} \frac{\partial F}{\partial p_3} \tag{9}$$

On the other hand, to obtain $\frac{d\theta}{d\tau}$, Qian and Leung (2004) gave the relationship between $dp_1, dp_3$ and $\tau$ as

$$\begin{aligned} \frac{dp_1}{d\tau} &= (p_1 \frac{\partial F}{\partial p_1} + p_3 \frac{\partial F}{\partial p_3})^{-1} \frac{\partial F}{\partial x} \\ \frac{dp_3}{d\tau} &= (p_1 \frac{\partial F}{\partial p_1} + p_3 \frac{\partial F}{\partial p_3})^{-1} \frac{\partial F}{\partial z} \end{aligned} \tag{10}$$

By obtaining the derivation of equation (7), we can also obtain the slowness parameter derivation directly:

$$\frac{dp_1}{d\tau} = \frac{v\cos\theta - \frac{\partial v}{\partial \theta}\sin\theta}{v^2}\frac{d\theta}{d\tau} - \frac{\sin\theta}{v^2}(\frac{\partial v}{\partial x}\frac{dx}{d\tau} + \frac{\partial v}{\partial z}\frac{dz}{d\tau})$$

$$\frac{dp_3}{d\tau} = \frac{v\sin\theta - \frac{\partial v}{\partial \theta}\cos\theta}{v^2}\frac{d\theta}{d\tau} - \frac{\cos\theta}{v^2}(\frac{\partial v}{\partial x}\frac{dx}{d\tau} + \frac{\partial v}{\partial z}\frac{dz}{d\tau})$$

(11)

Combining equations (10) with (11), we obtain:

$$\frac{d\theta}{d\tau} = (p_1\frac{\partial F}{\partial p_1} + p_3\frac{\partial F}{\partial p_3})^{-1}(v\frac{\partial F}{\partial x}\cos - v\frac{\partial F}{\partial z}\sin)$$

(12)

When the rays are nearly level, they are required to meet the following conditions (Symes and Qian, 2003):

$$\frac{dz}{d\tau} = (p_1\frac{\partial F}{\partial p_1} + p_3\frac{\partial F}{\partial p_3})^{-1}\frac{\partial F}{\partial p_3} > 0$$

(13)

At this time

$$\frac{dx}{dz} = \frac{\partial F}{\partial p_1}(\frac{\partial F}{\partial p_3})^{-1}$$

(14)

$$\frac{d\theta}{dz} = (\frac{\partial F}{\partial p_3})^{-1}(v\frac{\partial F}{\partial x}\cos - v\frac{\partial F}{\partial z}\sin\theta)$$

(15)

Equations (14) and (15) are used here as the anisotropic ray tracing and dynamic ray tracing equations in next section.

## 4. Ray tracing and dynamic ray tracing equations

In this work, ray tracing for the Eulerian Gaussian beam has been established based on level set theory. Generally, the level set equation has the form (Qian and Leung, 2004)

$$\frac{d\phi}{dz} = \phi_z + u\phi_x + w\phi_\theta = 0$$

(16)

in the equations

$$u = \frac{dx}{dz}$$

(17)

$$w = \frac{d\theta}{dz}$$

(18)

The derivatives (17) and (18) are Equations (14) and (15), respectively where above. According to Equation (16), we can obtain the partial differential equations with ray tracing

as follows:

$$\frac{dx}{dz} = x_z + ux_x + wx_\theta = 0$$
$$\frac{d\vartheta}{dz} = \vartheta_z + u\vartheta_x + w\vartheta_\theta = 0 \quad (19)$$
$$\frac{d\tau}{dz} = \tau_z + u\tau_x + w\tau_\theta = p_3 + p_1\frac{dx}{dz}$$

The initial conditions for numerical solution (Leung et al., 2007) are:

$$x(0, x, \theta) = x$$
$$\vartheta(0, x, \theta) = \theta \quad (20)$$
$$\tau(0, x, \theta) = 0$$

The dynamic ray tracing equations (Leung et al., 2007) are:

$$B_z + uB_x + wB_\theta = -H_{xp}B - H_{xx}C$$
$$C_z + uC_x + wC_\theta = -H_{pp}B + H_{xp}C \quad (21)$$

where

$$H_{pp} = vu_1^2$$
$$H_{xp} = \frac{u_1^2 v_x N^T}{v} \quad (22)$$
$$H_{xx} = \frac{u_1(vv_{xx} - 3v_x v_x^T) + u_1^3 v_x v_x^T}{v^3}$$

For isotropic medium $u_1 = \frac{1}{\cos\theta}$, For anisotropic media $u_1 = \cos\theta + \sin\theta\frac{dx}{dz}$. Thus

$$H_{pp} = v(\cos\theta + \sin\theta\frac{dx}{dz})^2$$
$$H_{xp} = \frac{(\cos\theta + \sin\theta\frac{dx}{dz})^2 v_x N^T}{v} \quad (23)$$
$$H_{xx} = \frac{(\cos\theta + \sin\theta\frac{dx}{dz})(vv_{xx} - 3v_x v_x^T) + (\cos\theta + \sin\theta\frac{dx}{dz})^3 v_x v_x^T}{v^3}$$

Similar to isotropic media, when the ray tracing equations are solved numerically, the initial value is amended as follows:

$$B(0,x,\theta) = i \ni \cos\theta + \sin\theta \frac{dx}{dz}$$
$$C(0,x,\theta) = \cos\theta + \sin\theta \frac{dx}{dz}$$
(24)

## 5. Eulerian Gaussian beam

For the Gaussian beam method, the complex travel time is treated with a second-order Taylor expansion (Leung et al, 2007):

$$\tau(z,x) = \tau(z,x_s) + \frac{1}{vu_1} N^T [x-X] + \frac{1}{2}(x-X)^T M(x-X) \tag{25}$$

The Hessian matrix is $M = BC^{-1}$, where $\tau$ is the travel time. The value of $B, C$ is obtained by solving the dynamic ray tracing equations, as described later in this paper; the direction of the central beam is $N: u_1 = \sqrt{1+N^2}$

The amplitude value of an arbitrary point on the ray can be expressed as follows (Leung et al., 2007):

$$A(z,x_s) = \sqrt{\frac{v(z,x) t_0(0,x_s) \det[C(0)]}{v(0,x_s) t_0(z,x_s) \det[C(z,x,\theta)]}} \tag{26}$$

Substituting Equations (25) and (26) into Equation (2), the Eulerian Gaussian beam in two-dimensional isotropic media has the form (Leung et al., 2007):

$$u_{beam} = \sqrt{\frac{v(z,x)}{v(0,x_s) \det[C(z,x,\theta)] \cos\theta}} \exp\{i\omega[\tau(z,x,\theta) + \frac{\sin\theta}{v}(x_0 - x) + \frac{1}{2} B(z,x,\theta) C^{-1}(z,x,\theta)](x_0 - x)^2\} \tag{27}$$

We can obtain the following equation for the travel time in anisotropic media:

$$\tau(z,x) = \tau(z,x,\theta) + \frac{\sqrt{\sin\theta\left[\sin\theta(\frac{dx}{dz})^2 + 2\cos\theta\frac{dx}{dz} - \sin\theta\right]}}{v(\cos\theta + \sin\theta\frac{dx}{dz})}(x_0 - x)$$
$$+ \frac{1}{2} B(z,x,\theta) C^{-1}(z,x,\theta)(x_0 - x)^2 \tag{28}$$

In the high-frequency approximation, by using the steepest descent method for stationary phase analysis (Bleistein, 1984), we obtain the first and the second derivative of travel time as follows:

$$T_{0,N}\bigg|_{v=(x-x_s)/z} = 0 \tag{29}$$

$$T_{0,NN}\bigg|_{v=(x-x_s)/z} = \left(\frac{-r}{v_0 a}\right)\frac{1}{(\cos\theta + \sin\theta\frac{dx}{dz})^4} \tag{30}$$

By using the stationary phase method, we obtain:

$$G(z,x) = \sqrt{\frac{2\pi}{\omega}} \frac{g_0 g_i g}{\sqrt{\det[T_{0,NN}]_k}} \exp(-i\omega T)$$

$$= \sqrt{\frac{2\pi}{\omega}} \frac{\Psi_0 g_i g_k}{(\cos\theta + \sin\theta\frac{dx}{dz})^2} \sqrt{\frac{1}{\det[C]\det[T_{0,NN}]}} \exp(-i\omega T) \tag{31}$$

where $g_i g_k$ are the eigenvectors corresponding the Christoffel matrix [Equation (4)]. The Green's function based on the general theory of high-frequency radiation in anisotropic media is (Červený, 2001):

$$G = g_i g \frac{1}{4\pi v_0 \sqrt{\det[C]}} \exp(-i\omega T) \tag{32}$$

We compared Equations (32) with (31) to obtain the weighting function in Eulerian Gaussian beam superposition:

$$\Psi_0 = \sqrt{\frac{\omega \det[T_{0,NN}]}{2\pi}} \frac{1}{4\pi v_0} (\cos\theta + \sin\theta\frac{dx}{dz})^2 \tag{33}$$

Therefore the Green's function based on Eulerian Gaussian beam superposition can be written as:

$$G = \Psi_0 \int_{(x,\theta)\in\Gamma(z,x_s)} \sqrt{\frac{v(z,x)}{v(0,x_s)\det[C(z,x,\theta)]}} (\cos\theta + \sin\theta\frac{dx}{dz}) d\vartheta \exp\{i\omega[\tau(z,x,\theta) + \frac{\sqrt{\sin\theta[\sin\theta(\frac{dx}{dz})^2 + 2\cos\theta\frac{dx}{dz} - \sin\theta]}}{v(\cos\theta + \sin\theta\frac{dx}{dz})}$$

$$(x_0 - x) + \frac{1}{2}B(z,x,\theta)C^{-1}(z,x,\theta)(x_0 - x)^2]\} \tag{34}$$

where the integration region is:

$$\Gamma\{(x,z,\theta): X(z,x,\theta) = x_s\} \tag{35}$$

## 6. Conclusions

The Gaussian beam method has an important role in high-frequency wave field calculations and seismic migration. However, the difficulties during the calculation process faced by the Gaussian beam method in high-frequency wave fields and seismic imaging have not yet been solved – that is, the time-consuming coordinate transformation and also the phase distribution. The Eulerian Gaussian beam method has been proposed to solve these two problems. We derived the equations for the anisotropic Eulerian Gaussian beam. The directional derivative of the ray parameters was obtained from the Hamiltonian form of the eikonal equation and the $x$, $z$ coordinates of the derivative of the travel time. We derived anisotropy travel time formulas based on the physical meaning of the direction factor $u_1$, as well as the relationship between the angle $\theta$ and $u_1$. In addition, we obtained the ray tracing and dynamic ray tracing equations. The Eulerian Gaussian beam method is within the scope of ray theory and therefore we used the stationary phase method to obtain the weighting function $\Psi_0$ and the changes in the weighting function with the source point of the travel time. In future studies, we plan to develop numerical algorithms for the high-frequency wave field and seismic migration using the anisotropic Eulerian Gaussian beam.